%Paper: 9109022
%From: SEN%TIFRVAX.BITNET@cunyvm.cuny.edu
%Date: Thu, 12 Sep 91 15:11 IST

\input phyzzx

{}~\hfill\vbox{\hbox{TIFR/TH/91-39}\hbox{September, 1991}}

\NPrefs
\def\define#1#2\par{\def#1{\Ref#1{#2}\edef#1{\noexpand\refmark{#1}}}}
\def\con#1#2\noc{\let\?=\Ref\let\<=\refmark\let\Ref=\REFS
         \let\refmark=\undefined#1\let\Ref=\REFSCON#2
         \let\Ref=\?\let\refmark=\<\refsend}

\define\ANTO
R.C. Myers, Phys. Lett. {\bf B199} (1988) 371;
I. Antoniadis, C. Bachas, J. Ellis and D. Nanopoulos, Phys. Lett. {\bf
B211} (1988) 393, Nucl. Phys. {\bf
B328} (1989) 117;
S.P. de Alwis, J. Polchinski and R. Schimmrigk, Phys. Lett. {\bf B218}
(1989) 449.

\define\SCHUBERT
C. Schubert, MIT preprint CTP 1977.

\define\CONE
R. Dijkgraaf, E. Verlinde and H. Verlinde, Comm. Math. Phys. {\bf 115}
(1988) 649.

\define\BPZ
A.~Belavin, A.~M.~Polyakov and A.~B.~Zamolodchikov, Nucl.~Phys. {\bf B241}
(1984) 333.

\define\SENPOLY
A.~Sen,  Phys.~Lett. {\bf B241} (1990) 350.

\define\KAKU
M. Kaku and J. Lykken, Phys. Rev. {\bf D38} (1988) 3067;
M. Kaku, preprints CCNY-HEP-89-6, Osaka-OU-HET 121.

\define\BACKGROUND
A.~Sen, Nucl.~Phys. {\bf B345} (1990) 551.

\define\NONPOL
M.~Saadi and B.~Zwiebach, Ann.~Phys. {\bf 192} (1989) 213;
T.~Kugo, H.~Kunitomo, and K.~Suehiro, Phys.~Lett. {\bf 226B} (1989) 48;

\define\GAUGEINV
T.~Kugo and K.~Suehiro, Nucl.~Phys. {\bf B337} (1990) 434.

\define\BACKTWO
A.~Sen, Nucl.~Phys. {\bf B347} (1990) 270.

\define\RENORM
A.~Sen, Phys.~Lett. {\bf B252} (1990) 566;

\define\SYCL
S.~Mukherji and A.~Sen, preprint TIFR/TH/91-12 (to appear in Nucl. Phys.
B)

\define\NULL
S. Mukherji, S. Mukhi and A. Sen, preprint TIFR/TH/91-25 (to appear in
Phys. Lett. B).

\define\BLACK
S. Mukherji, S. Mukhi and A. Sen, preprint TIFR/TH/91-28.

\define\WITTEN
E. Witten,     preprint IASSNS-HEP-91/12.

\define\MANDAL
G. Mandal, A.M. Sengupta and S.R. Wadia, preprint IASSNS-HEP-91/10.

\define\ROCEK
M. Rocek, K. Schoutens and A. Sevrin, preprint IASSNS-HEP-91/14.

\define\BARDACKI
K. Bardacki, M. Crescimannu, and E. Rabinovici, Nucl.
Phys. {\bf B344} (1990) 344.

\define\DVV
R. Dijkgraaf, E. Verlinde and H. Verlinde, preprint PUPT-1252,
IASSNS-HEP-91/22.

\define\EFR
S. Elitzur, A. Forge and E. Rabinovici, Preprint RI-143-90.

\define\BANE
I. Bars and D. Nemeschansky, Nucl. Phys. {\bf B348} (1991) 89.

\define\SOZW
H. Sonoda and B. Zwiebach, Nucl. Phys. {\bf B331} (1990) 592.

\define\HLOOP
H.~Hata, Phys.~Lett. {\bf 217B} (1989) 438, 445; Nucl.~Phys. {\bf B329}
(1990) 698; {\bf B339} (1990) 663.

\define\SAADI
M. Saadi, Mod. Phys. Lett. {\bf A5} (551) 1990; Int. J. Mod. Phys. {\bf
A6} (1991) 1003.

\define\ZWIEBACH
B.~Zwiebach, Mod.~Phys.~Lett. {\bf A5} (1990) 2753;
Mod.~Phys.~Lett. {\bf A5} (1990) 2753;
Phys. Lett. {\bf B241} (1990) 343; Comm. Math. Phys. {\bf 136} (1991) 83.

\define\ZWIEOPEN
B. Zwiebach, Phys. Lett. {\bf B256} (1991) 22; preprint MIT-CTP-1908.

\define\MKAKU
L. Hua and M. Kaku, Phys. Lett. {\bf B250} (1990) 56;
M. Kaku, Phys. Lett. {\bf B250} (1990) 64.

\def\BLACKHOLE{\con\EFR\MANDAL\WITTEN\ROCEK\BARDACKI\DVV\BANE\noc}
\def\OTHERS{\con\KAKU\SOZW\HLOOP\SAADI\ZWIEBACH\MKAKU\ZWIEOPEN\SCHUBERT\noc}

\def\vp{\varphi}
\def\vpzz{\vp(z,\bz)}
\def\vpizz{\varphi_i(z,\bz)}
\def\lo{\lambda^{(0)}}
\def\hs{\hat S}
\def\hqb{\hat Q_B}
\def\hp{\hat\Psi}
\def\tp{\tilde\Psi}
\def\ts{\tilde S}
\def\tqb{\tilde Q_B}
\def\pcl{\Psi_{cl}}
\def\bz{{\bar z}}
\def\c{\circ}
\def\cs{{\cal S}}
\def\l{\langle}
\def\r{\rangle}
\def\ch{{\cal H}}
\def\p{\partial}
\def\bb{\bar b}
\def\bc{\bar c}

\title{SOME APPLICATIONS OF STRING FIELD THEORY\foot{Invited talk given at
the `Strings and Symmetries, 1991' conference, held at Stony Brook, May
25-30, 1991}}

\author{Ashoke Sen}

\address{Tata Institute of Fundamental Research, Homi Bhabha Road, Bombay
400005, India}
\centerline{e-mail address: sen@tifrvax.bitnet}
\vskip .2in

\abstract

We study general properties of the classical solutions in non-polynomial
closed string field theory and their relationship with two dimensional
conformal field theories.
In particular we discuss how different conformal field theories which are
related by marginal or nearly marginal deformations can be regarded as
different classical solutions of some underlying string field theory.
We also discuss construction of a classical solution labelled by infinite
number of parameters in string field theory in two dimensions.
For a specific set of values of the parameters the solution can be
identified to the black hole solution.

\endpage

In this talk I shall discuss the relationship between the
classical
solutions of the equations of motion in string field theory and two
dimensional conformally invariant field theories.
Let us begin by giving a brief motivation for this study.
Let us consider two different conformal field theories, each with central
charge 26. Thus both of them form consistent background for the
formulation of bosonic string theory.
A question which naturally arises at this stage is, can these two
different conformal field theories be considered as two different
classical
solutions of some underlying string field theory?
The same question may be put in another way.
Let us consider formulating string field theory around the background of
one of the conformal field theories, so that the S-matrix elements
calculated from this string field theory agree with those calculated from
the first quantized Polyakov prescription.
The question then would be: can we now identify a specific classical
solution of this string field
theory which will represent string theory formulated around the second
conformal field theory?
In this talk I shall address the question in this second form.

Yet another way of asking the question is as follows.
Let S denote the string field theory action formulated in the background
of the first conformal field theory and $S'$ the string field theory
action formulated around the background of the second conformal field
theory.
One then asks: is there a field redefinition which will produce the action
of the second conformal field theory starting from the first one?
In this formulation the question may also be regarded as the question of
background independence of string field theory, since an answer to the
above question in the affirmative will guarantee that the string field
theory is intrinsically independent of the background in which it is
formulated.

The plan of the talk will be as follows.
First I shall show that any classical solution of the string field theory
corresponds, in some sense, to a generalized conformal field theory.
(The precise sense in which it happens will be discussed later.)
Next I shall show how, given a marginal deformation of the original
conformal field theory around which string field theory is formulated, one
can construct a solution of the string field theory equations of motion
representing the new conformal field theory obtained by this marginal
deformation.
I shall also show how, using the techniques of string field theory, one
can construct an infinite parameter solution in the string theory
describing $c=1$ conformal field theory coupled to gravity.
Finally I shall consider a case where the original conformal field theory
has a nearly marginal deformation which gives rise to a new conformal
field theory.
I shall again show how to construct a solution of the string field theory
equations of motion that will represent this new conformal field theory.
(As we shall see, the situation here is somewhat more subtle since the new
conformal field theory does not have the same central charge as the
original conformal field theory.
Hence we need to combine this with a second conformal field theory which
will compensate for the change in the central charge.)

Let us start with a brief review of the formulation of non-polynomial
closed string field theory\NONPOL\GAUGEINV\ in the background of a given
conformal field theory.
(For related work in non-polynomial closed string field theory, see
refs.\OTHERS.)
As is well known, the first quantized string theory is based on a two
dimensional action given by,
$$
\cs=\cs_{matter}+\cs_{ghost}
\eqn\eone
$$
where $\cs_{matter}$ represents a conformal field theory of central charge
26, and $\cs_{ghost}$ represents the action for the ghost fields given by,
$$
\cs_{ghost}=\int d^2z (b\p_\bz c+\bb\p_z \bc)
\eqn\etwo
$$
Let $\ch$ denote the combined hilbert space of the matter and ghost
theory.
This Hilbert space is characterized by the existence of a unique SL(2,C)
vacuum $|0\r$ which is annihilated by the Virasoro generators $L_n$ for
$n\ge -1$.
Here $L_n=L^M_n+L^G_n$, where $L^M_n$ and $L^G_n$ denote the
Virasoro generators of the
matter and ghost sectors respectively.
Using the usual rules of conformal field theory\BPZ, we  can then
establish a one to one correspondence between the states $|\Phi\r$ in
$\ch$
and the local operators $\Phi(z,\bz)$ in the theory through the relation:
$$
|\Phi\r=\Phi(0)|0\r
\eqn\ethree
$$
We shall now define ghost number of various operators in the theory by
assigning the fields $b$, $\bb$ to have ghost number $-1$, $c$, $\bc$ to
have ghost number $1$, and all the matter sector fields to have ghost
number
0.
This uniquely defines the ghost numbers of all the operators in the
theory.
We shall further assign the vacuum state $|0\r$ to have ghost number 0.
This then defines the ghost number of all the states in the theory:
the ghost number of the state $|\Phi\r$ given in eq.\ethree\ is the same
as that of the operator $\Phi(z,\bz)$.
Finally we define the BRST charge of the first quantized theory as,
$$
Q_B=\sum_{n=-\infty}^\infty c_{-n} L^M_n -{1\over 2} \sum_{m,n
=-\infty}^\infty (m-n) :b_{m+n}c_{-m}c_{-n}:
\eqn\efour
$$
where $c_n$ etc. are defined through the relations:
$$
c(z)=\sum_nc_nz^{-n+1},~~\bc_n=\sum_n \bc_n\bz^{-n+1},
{}~~b(z)=\sum_n b_n z^{-n-2},~~\bb(\bz)=\sum_n \bb_n\bz^{-n-2}
\eqn\efive
$$
The BRST charge is nilpotent, i.e. $(Q_B)^2=0$.

A first quantized string state is an element of the Hilbert space $\ch$.
More precisely, the physical states of the first quantized theory are the
BRST invariant states of $\ch$, with the equivalence relation that
two states are physically equivalent if they differ by a BRST exact
state.
Thus the naive guess for the space of second quantized string field (which
should correspond to the wave-function or the state of the first quantized
string
theory) would
be the whole Hilbert space $\ch$, with BRST invariance being the on-shell
condition, and equivalence of physical states differing by a BRST exact
state being a consequence of gauge invariance of the string field theory.
It turns out, however, that this naive expectation is wrong, at least in
the current formulation of string field theory.
First we define a subspace $\ch'$ of $\ch$ as the space of
all states $|\Phi\r$ satisfying:
$$c_0^-|\Phi\r=0,~~~~~L_0^-|\Phi\r=0
\eqn\esix
$$
where $c_0^-=(c_0-\bc_0)/\sqrt 2$, $L_0^-=(L_0-\bar L_0)/\sqrt 2$.
An off-shell string field configuration $|\Psi\r$ is then taken to be an
arbitrary state in $\ch'$ of ghost number 3.\foot{Equivalently we could
call $b_0^-|\Psi\r$ our string field.
This has ghost number 2 and is annihilated by $b_0^-$.
Normally what we call first quantized string state is $b_0^-|\Psi\r$.}

In order to specify a string field theory, we must write down an
expression for the action in terms of the string field $|\Psi\r$.
This is done in the following way.
We first introduce, for all values of $N\ge 3$, a multilinear map (denoted
by $\{A_1\ldots A_N\}$) from the $N$
fold tensor product of the Hilbert space $\ch$
to the space of complex numbers, and another multilinear
map (denoted by $[A_1\ldots A_{N-1}]$) from the $N-1$ fold tensor product
of the Hilbert space $\ch$ to
the Hilbert space $\ch$, satisfying the following properties:
$$\eqalign{
&\{A_1\ldots A_N\}=(-1)^{n_1+1}\l A_1|[A_2\ldots A_N]\r\cr
&\{A_1 A_2\ldots (\alpha A_i+\beta B_i)\ldots A_N\}=\alpha\{A_1\ldots
A_i\ldots A_N\} +\beta\{A_1\ldots B_i\ldots A_N\}\cr
&b_0^-[A_1\ldots A_{N-1}]=L_0^-[A_1\ldots A_{N-1}]=0\cr
&\{A_1\ldots A_N\}=(-1)^{(n_i+1)(n_{i+1}+1)}\{A_1\ldots A_{i-1}A_{i+1} A_i
A_{i+2}\ldots A_N\}\cr
&Q_B[A_1\ldots A_{N-1}]=\sum_{i=1}^{N-1}(-1)^{\sum_{j=1}^{i-1}(n_j+1)}
[A_1\ldots Q_BA_i\ldots A_N]\cr
& \quad\quad\quad
-\sum_{\{i_l,j_k\}\atop l\ge 1, k\ge 2}(-1)^{\sigma\{i_l,j_k\}} [A_{i_1}
\ldots A_{i_l}c_0^-[A_{j_1}\ldots A_{j_k}]]\cr
}
\eqn\eseven
$$
where $b_0^-=(b_0-\bb_0)/\sqrt 2$, $n_i$ denotes the ghost number of the
state $A_i$, and $\sigma(\{i_l,j_k\})$ is a phase factor which is the
phase picked up during the rearrangement of the operators $Q_B,
b_0^-A_i,\ldots b_0^- A_N$ to $b_0^-A_{i_1},\ldots b_0^-A_{i_l}, Q_B,
b_0^-A_{j_1},\ldots b_0^-A_{j_k}$.

We shall discuss the explicit construction of these multilinear  maps
later.
For the time being let us assume that a set of multilinear maps satisfying
eq.\eseven\ exists, and see how this helps us formulate a gauge invariant
string field theory.
In fact, the action of string field theory takes a simple form when
expressed in terms of these maps:
$$
S(\Psi)={1\over 2}\l\Psi|Q_B b_0^-|\Psi\r+\sum_{N=3}^\infty {g^{N-2}\over
N!} \{\Psi^N\}
\eqn\eeight
$$
where $\{\Psi^N\}=\{\Psi\Psi\ldots \Psi\}$, and $Q_B$ is the BRST charge
of the first quantized string theory.
This action is invariant under the following gauge transformations:
$$
b_0^-\delta|\Psi\r=Q_Bb_0^-|\Lambda\r+\sum_{N=3}^\infty {g^{N-2}\over
(N-2)!} [\Psi^{N-2}\Lambda]
\eqn\enine
$$
where $|\Lambda\r$ is an arbitrary state in $\ch'$ with ghost number 2.
The proof of invariance of the action \eeight\ under the gauge
transformation \enine\ follows directly from the properties of $[~]$ and
$\{~\}$ listed in eq.\eseven\ and the nilpotence of the BRST charge $Q_B$;
we never need the explicit forms of the maps $[~]$ and $\{~\}$.
Note that the linearized equations of motion $Q_B b_0^-|\Psi\r=0$
implies that physical (on-shell) states correspond to BRST invariant states;
$-$ a result that matches with that of the first quantized  theory.
Furthermore,
the linearized gauge transformation law implies that two
on-shell states which differ by a BRST exact state are gauge equivalent,
again as expected from first quantized theory.

An explicit construction of the maps $\{~\}$ and $[~]$ satisfying the
properties listed in eq.\eseven\ was constructed in
refs.\NONPOL\GAUGEINV\ and can be expressed in terms of appropriate
correlation functions in
the combined matter-ghost conformal field theory\GAUGEINV\SENPOLY.
We shall not give the general construction here since it will not be
needed for our purpose, but shall illustrate the construction by giving
the expression for $\{A_1A_2A_3\}$.
This is constructed as follows.
Let $\tilde A_i$ be the local fields such that,
$$
\tilde A_i(0)|0\r=b_0^-|A_i\r
\eqn\eten
$$
Then,
$$
\{ A_1 A_2 A_3\}=-\l f_1\c \tilde A_1(0) f_2\c \tilde A_2(0) f_3\c\tilde
A_3(0) \r
\eqn\eeleven
$$
where $f_i$ ($1\le i\le 3$) are three conformal maps given by,
$$
f_1(z)=\bigg({1-iz\over 1+iz}\bigg)^{2/3},~~f_2(z)=e^{2\pi i/3} f_1(z),~~
f_3(z)=e^{4\pi i/3} f_1(z)
\eqn\etwelve
$$
and $f_i\c \tilde A_i(z,\bz)$ denotes the conformal transform of the field
$\tilde A_i(z,\bz)$ under the map $f_i$.
$\l~~\r$ denotes the correlation function in the combined matter-ghost
conformal field theory.
Thus we see that given three states $|A_1\r$, $|A_2\r$ and $|A_3\r$, the
quantity $\{A_1A_2A_3\}$ is completely defined by eqs.\eten-\etwelve.
The first of eq.\eseven\ also determines the state $[A_1A_2]$ completely
once $\{A_1A_2A_3\}$ is known for all states $A_i$.

Similar expressions exist for higher point functions $\{A_1\ldots A_N\}$,
such that they satisfy the relations given in eq.\eseven.
Furthermore, when one constructs a string field theory action based on
these maps and eq.\eeight, and computes the S-matrix elements after
appropriate gauge fixing, one can recover the usual expression for the
S-matrix given by the first quantized Polyakov prescription.
In fact the same identities \eseven\ which are responsible for the gauge
invariance of the theory also turns out to be crucial in obtaining the
correct S-matrix from string field theory.

Let us now investigate some properties of the equations of motion of the
theory.
The equations of motion obtained by varying the action \eeight\ with
respect to the string fields is given by:
$$
Q_B b_0^-|\Psi\r+\sum_{N=3}^\infty {g^{N-2}\over (N-1)!}[(\Psi)^{N-1}]=0
\eqn\ethirteen
$$
Let $\pcl$ be any classical solution of the equations of motion.
The question we shall first ask is: how does string field theory
formulated around
$\pcl$ look like\SENPOLY?
In particular, we would like to know if it looks like string field theory
formulated around a new conformal field theory.
In order to study this question, let us define shifted field:
$$
\hp=\Psi-\pcl
\eqn\efourteen
$$
and expand the action in terms of the field $\hp$.
It can be easily seen that the action takes the form:
$$
S(\Psi)=S(\pcl)+{1\over 2}\l\hp |\hqb b_0^-|\hp\r+\sum_{N=3}^\infty
{g^{N-2}\over N!}\{\hp^N\}'\equiv S(\pcl)+\hat S(\hp)
\eqn\efifteen
$$
where $\hqb$ is a new linear operator and $\{~~\}'$ is a new multilinear
map, defined as:
$$
\hqb b_0^-|A\r=Q_B b_0^-|A\r +\sum_{M=3}^\infty {g^{M-2}\over (M-2)!}
[(\pcl)^{M-2} A]
\eqn\esixteen
$$
$$
\{A_1\ldots A_N\}'=\{A_1\ldots A_N\}+\sum_{M=3}^\infty {g^{M-2}\over
(M-2)!} \{(\pcl)^{M-2}A_1\ldots A_N\}
\eqn\eseventeen
$$
Let us also define,
$$
[A_1\ldots A_{N-1}]'=[A_1\ldots A_{N-1}]+\sum_{M=3}^\infty {g^{M-2}\over
(M-2)!} [(\pcl)^{M-2}A_1\ldots A_{N-1}]
\eqn\eeighteen
$$
Note that the shifted action $\hat S(\hp)$ has the same form as the action
\eeight\ with $\Psi$ replaced by $\hp$, $Q_B$ replaced by $\hqb$, and
$\{~\}$ replaced by $\{~\}'$.
Furthermore, using the fact that $\pcl$ satisfies the equations of
motion \ethirteen, one can verify\SENPOLY\ the following relations
satisfied by
$\hqb$, $\{~\}'$ and $[~]'$:
$$\eqalign{
&(\hqb)^2=0\cr
&\{A_1\ldots A_N\}'=(-1)^{n_1+1}\l A_1|[A_2\ldots A_N]'\r\cr
&\{A_1 A_2\ldots (\alpha A_i+\beta B_i)\ldots A_N\}'=\alpha\{A_1\ldots
A_i\ldots A_N\}' +\beta\{A_1\ldots B_i\ldots A_N\}'\cr
&b_0^-[A_1\ldots A_{N-1}]'=L_0^-[A_1\ldots A_{N-1}]'=0\cr
&\{A_1\ldots A_N\}'=(-1)^{(n_i+1)(n_{i+1}+1)}\{A_1\ldots A_{i-1}A_{i+1} A_i
A_{i+2}\ldots A_N\}'\cr
&\hqb [A_1\ldots A_{N-1}]'=\sum_{i=1}^{N-1}(-1)^{\sum_{j=1}^{i-1}(n_j+1)}
[A_1\ldots \hqb A_i\ldots A_N]'\cr
& \quad\quad\quad
-\sum_{\{i_l,j_k\}\atop l\ge 1, k\ge 2}(-1)^{\sigma\{i_l,j_k\}} [A_{i_1}
\ldots A_{i_l}c_0^-[A_{j_1}\ldots A_{j_k}]']'\cr
}
\eqn\enineteen
$$
The first of these equations tell us that the operator $\hqb$ is
nilpotent.
The rest of the equations are identical to those in eq.\eseven\ with
$Q_B$, $\{~\}$ and $[~]$ replaced by $\hqb$, $\{~\}'$ and $[~]'$
respectively.

Note that the nilpotence of the original BRST charge $Q_B$ as well as the
set of relations given in eq.\eseven\ were consequences of conformal
invariance of the original system.
What we see here is that for every classical solution of string field
theory equations of motion we can construct a nilpotent operator $\hqb$
and a set of multilinear maps $\{~\}'$, $[~]'$ satisfying the same set of
properties given in eq.\eseven.
This leads us to believe that every classical solution of string field
theory corresponds to a conformal field theory or some generalization of
it, characterized by a nilpotent BRST charge $\hqb$ and the set of
multilinear maps $\{~\}'$ and $[~]'$.
It remains to be seen whether every classical solution of string field
theory indeed corresponds to a conformal field theory.
For this one needs to construct the new Virasoro generators $\hat L_n$,
$\hat{\bar L_n}$ corresponding to the classical solution $\pcl$.

Next let us turn towards construction of explicit solution of equations of
motion in string field theory.
It is known that given a conformal field theory with a marginal or nearly
marginal deformation, we can construct other nearby conformal field
theories by perturbing the original conformal field theory by the marginal
or nearly marginal operator.
The question we shall address is: given a string field theory constructed
in the background of the original conformal field theory, can we construct
explicit solutions of the string field theory equations of motion which
represent string theory formulated around the perturbed conformal field
theory?

Let us start from marginal perturbation.
Let $\cs_{CFT}$ be the two dimensional action of the original conformal
field theory, and $\vpizz$ be the set of all marginal operators in the
theory.
The action of the perturbed conformal field theory is then given by,
$$
\cs'_{CFT}=\cs_{CFT}-\lo_i\int d^2 z\vpizz+O((\lo)^2)
\eqn\etwenty
$$
where $\lo_i$ are small parameters.
To order $\lo$ we can indeed construct a solution of the string field
theory equations of motion.
This is given by:
$$
b_0^-|\pcl\r=\lambda_i|\varphi_i\r_{matter}\otimes c_1\bc_1|0\r_{ghost}
\eqn\etwentyone
$$
where $\lambda_i\propto\lambda_i^{(0)}$ are also small parameters, and
$|\varphi_i\r=\varphi_i(0)|0\r$.
The exact relation between $\lo_i$ and $\lambda_i$ will be determined
later.
Note that \etwentyone\ is a solution of the linearised equations
of motion since $b_0^-|\pcl\r$ is a BRST invariant state.

The question that we shall now pose is the following\SYCL?
Under what condition can we add systematic corrections to the solution
\etwentyone\ which gives us a solution of the complete string field theory
equations of motion to all orders in $\lambda_i$?
In order to study this question, let us take a general solution of the
form
$$
|\pcl\r=\lambda_i|\varphi_i\r_{matter}\otimes c_0^-c_1\bc_1|0\r_{ghost}
+\sum_n\lambda^n|\chi_n\r
\eqn\etwentytwo
$$
where $\lambda$ stands for any of the $\lambda_i$'s.
Let us also define $|\Psi_N\r=\sum_{n=1}^N\lambda^n|\chi_n\r$.
The equation of motion to $N$th order then takes the form:
$$
|E_N\r\equiv
Q_B b_0^-|\Psi_N\r+\sum_{M=3}^\infty {g^{M-2}\over (M-1)!}
[(\Psi_N)^{M-1}]
= O(\lambda^{N+1})
\eqn\etwentytwo
$$
We shall try to prove the existence of a solution of the above equation by
induction.
In other words we shall assume that eq.\etwentytwo\ holds for a given
value of $N$ and then try to show that under certain conditions it also
holds for $N$ replaced by $N+1$.
Since the equation holds for $N=1$, this would then imply that it holds
for all values of $N$.

The equation $|E_{N+1}\r=O(\lambda^{N+2})$ can be rewritten as,
$$
\lambda^{N+1} Q_B b_0^-|\chi_{N+1}\r=-Q_B b_0^-|\Psi_N\r
-\sum_{M=3}^\infty {g^{M-2}\over (M-1)!} [(\Psi_N)^{M-1}]
+O(\lambda^{N+2})
\eqn\etwentythree
$$
where on the right hand side inside $[~]$ we have replaced $\Psi_{N+1}$ by
$\Psi_N$.
This is allowed since the difference between $[(\Psi_N)^{M-1}]$ and
$[(\Psi_{N+1})^{M-1}]$ can easily be seen to be of order $\lambda^{N+2}$
for $M\ge 3$.
We need to show that it is possible to adjust $|\chi_{N+1}\r$ so that
eq.\etwentythree\ is satisfied.
The non-triviality in this problem comes from the fact that $Q_B
b_0^-|\chi_{N+1}\r$ is a pure gauge state; hence unless the state
appearing on the right hand side of the equation is also pure gauge, this
equation cannot be satisfied by adjusting $|\chi_{N+1}\r$.
We shall now determine the condition under which the right hand side of
eq.\etwentythree\ is pure gauge.

We rewrite eq.\etwentythree\ as,
$$\eqalign{
\lambda^{N+1}\l s| Q_B b_0^-|\chi_{N+1}\r=&-\l s|Q_B b_0^-|\Psi_N\r
-\sum_{M=3}^\infty {g^{M-2}\over (M-1)!} \l s|[(\Psi_N)^{M-1}]\r
+O(\lambda^{N+2})\cr &{\rm for~all~}\l s|\cr
}
\eqn\etwentyfour
$$
The set of all states $\l s|$ (or, equivalently, $\l s |b_0^-$) may be
divided into three classes: pure
gauge states for which $\l s|b_0^-=\l\Lambda|b_0^-Q_B$ for some
$\l\Lambda|$; physical states which are not pure gauge, but satisfy
$\l s|b_0^-Q_B=0$; and unphysical states for which $\l s|b_0^-Q_B\ne 0$.
The statement of the preceeding paragraph may now be translated into a
different language.
If $\l s| b_0^-$ is either a physical or a pure gauge state, then the
inner
product of $\l s|$ with $Q_B b_0^-|\chi_{N+1}\r$ vanishes automatically.
Thus in order to satisfy eq.\etwentyfour\ for such states $\l s|$, we must
make sure that the right hand side of the equation vanishes identically
when the state $\l s| b_0^-$ is either physical or pure gauge.
On the other hand, if the state $\l s |b_0^-$ is unphysical, then $\l s|$
has non-vanishing inner product
with $Q_B b_0^-|\chi_{N+1}\r$.
Hence for such states eq.\etwentyfour\ can be satisfied by
adjusting $|\chi_{N+1}\r$.
(See
ref.\SYCL\ for a detailed proof of this).

We start with the case where $\l s|$ is a pure gauge state.
In this case eq.\etwentyfour\ takes the form:
$$
\sum_{M=3}^\infty {g^{M-2}\over (M-1)!} \l \Lambda| Q_B|[(\Psi_N)^{M-1}]\r
= O(\lambda^{N+2})~~~~{\rm for~all~}\l \Lambda|
\eqn\etwentyfive
$$
In deriving eq.\etwentyfive\ we have used various properties of
$[(\Psi)^{M-1}]$ listed in eq.\eseven\ to express it as
$b_0^-c_0^-[(\Psi)^{M-1}]$, and also that
$\{Q_B,b_0^-\}c_0^-|[(\Psi)^{M-1}]\r
=L_0^-c_0^- |[(\Psi)^{M-1}]\r=0$.
We may evaluate the left hand side of eq.\etwentyfive\ using the
properties of $Q_B |[(\Psi)^{M-1}]\r$ given in eq.\eseven.
Using the equation of motion \etwentytwo\ to order $\lambda^N$ it can be easily
seen that the expression given in eq.\etwentyfive\ is of order
$\lambda^{N+2}$\SYCL.
This, in turn, shows that the condition $\l s|E_{N+1}\r=O(\lambda^{N+2})$
is satisfied automatically if $\l s|$ is a pure gauge state.

If, on the other hand, we choose the state $\l s|$ to be a physical state,
then the left hand side of eq.\etwentyfour\ still vanishes identically,
but there is no reason for the right hand side to vanish.
Thus these equations represent genuine obstructions to extending the
lowest order solution \etwentyone\ to higher orders.
To understand the significance of these equations, let us study the order
$\lambda^2$ contribution to these equations.
For this we need to substitute the lowest order solution \etwentyone\ in
the right hand side of eq.\etwentyfour, and take the inner product of this
equation with any of the states $ \l s_i| = \l\vp_i| c_{-1}\bar c_{-1}
c_0^-$ for which $\l s_i|b_0^-$ is physical.
This can be easily evaluated and we get the set of equations:
$$
\sum_{j,k} C_{ijk}\lambda_j\lambda_k = 0
\eqn\etwentysix
$$
where $C_{ijk}$ is the coefficient of $\vp_i$ in the operator product of
$\vp_j$ with $\vp_k$.
One can now easily recognise the left hand side of the above equation as
the $\beta$-function of the perturbed conformal field theory.
Thus we see that the obstructions represented by taking the inner product
of eq.\etwentythree\ with physical states are string field theoretic
generalization of the $\beta$-functions.
In order to get a non-trivial solution of the classical equations of
motion in string field theory representing a perturbed conformal field
theory, one must demand that the $\beta$-function of the perturbed theory
must vanish.

Before we go on, we shall apply the procedure outline above to construct a
set of classical solutions in string field theory representing $c=1$
conformal field theory coupled to gravity\BLACK.
The corresponding conformal field theory is given by the direct sum of two
theories, a free scalar ($\phi_M$) field theory with central charge 1, and
the Liouville field ($\phi_L$) theory which we shall represent by a scalar
field theory with a background charge $Q$ at infinity.
$Q$ is adjusted so that the
central charge $1+ 3Q^2$ of this theory is equal to 25.
The field $\phi_L$ has Euclidean signature, whereas the matter field
$\phi_M$ may have Euclidean or Minkowski signature.
We shall refer to $\phi_M$ as `time' and $\phi_L$ as space from now on.
The region $\phi_L\to-\infty$ corresponds to weak coupling region, hence
we shall refer to it as the asymptotic region.

We shall now try to construct `static', asymptotically flat solutions in
this theory.
In other words we shall look for solutions which carry zero $\phi_M$
momentum, and Liouville momentum of the form $- i k_L$ with $k_L>0$, so
that the corresponding background $e^{k_L\phi_L}$ decays as
$\phi_L\to-\infty$.
The first step is to classify the physical states with this property.
It is well known\CONE\ that in the zero momentum sector, the primary
states in
the $c=1$ conformal field theory are labelled by two integers $r$ and $s$
and has dimension $(r^2, s^2)$.
Let us denote these states by $|\vp_{r,s}\r$.
It turns out that physical states (of ghost number 2) in the combined
matter, Liouville, ghost
theory satisfying the conditions of being static and asymptotically flat
are of the form\NULL\ $|\vp_{r,r}\r_M\otimes |k_L = (r+1)Q/\sqrt 2\r_L
\otimes c_1\bc_1|0\r_{ghost}$.
Thus to lowest order a solution of the classical equations of motion may
be written as,
$$
b_0^-|\pcl\r = \sum_r \lambda_r |\vp_{r,r}\r_M\otimes |k_L =(r+1)Q/\sqrt
2\r_L\otimes c_1\bc_1|0\r_{ghost}
\eqn\etwentyseven
$$
The question now is whether higher order corrections cause any obstruction
to extending these solutions to all orders in $\lambda_r$.
For this we only need to look for the inner product of the equations of
motion with the physical states.
Using the rule of addition of Liouville momentum, and the fusion rules of
$c=1$ conformal field theory, one can show\BLACK\ that the inner product
of the
equation of motion with any physical state vanishes identically in this
case.
Thus the solution exists to all orders in $\lambda_r$.
This establishes the existence of a solution in string field theory
labelled by infinite number of parameters.

Note that in this case, the expansion of the solution in powers of
$\lambda_r$ may also be regarded as an expansion in powers of
$e^{Q\phi_L}$, since each $\lambda_r$ comes with a specific power of
$e^{Q\phi_L}$ in the solution.
The expansion is good for large nagative $\phi_L$, but breaks down for
large positive $\phi_L$.
One may be able to analytically continue the solution past the point where
the expansion diverges, or
the solution may have a genuine singularity there.
With our present technology of string field theory, it is not possible to
answer which of these possibilities is correct.

Also note that if we choose $\lambda_1\ne 0$, but $\lambda_r =0$ for $r\ne
1$, then the asymptotic expansion of the solution coincides with the
recently discovered black hole solution in $c=1$ string theory\BLACKHOLE.
It can be shown\SYCL\ that given the asymptotic expansion, the solution in
string field theory is uniquely determined up to gauge transformations.
Thus we can say that for these values of the parameters, the solution does
represent the black hole solution.

Let us now turn to the case where the original conformal field theory has
a nearly marginal operator.
Let $\cs_{CFT}$ be the action of the original conformal field theory, and
$\vpzz$ be the nearly marginal operator of dimension $(1-y, 1-y)$.
It is known that in this case, if the operator product of $\vp$ with
itself an arbitrary number of times do not produce any other (nearly)
marginal operator except $\vp$ itself, then the perturbed action:
$$
\cs' = \cs_{CFT} - \lo \int d^2 z\vpzz
\eqn\etwentyeight
$$
gives a conformal field theory CFT$'$ for:
$$
\lo = {2 y\over C_{\vp\vp\vp}} +O(y^2)
\eqn\etwentynine
$$
where $C_{\vp\vp\vp}$ is the coefficient of $\vp$ in the $\vp\vp$ operator
product expansion.
The central charge of the perturbed conformal field theory is given by,
$$
c'_{CFT} = c_{CFT} - {8 y^3\over (C_{\vp\vp\vp})^2} + O(y^4)
\eqn\ethirty
$$
{}From this we see that there is an immediate problem in generalising the
previous analysis to this case: if the original conformal field theory
describes a consistent background for the formulation of string theory,
and hence has central charge 26, then the new conformal field theory does
not.
Thus we need to be somewhat more careful in formulating the problem.
For this let us consider string theory formulated around the background of
the original conformal field theory $CFT$ and a free scalar field ($X^0$)
theory with background charge $Q$, so that the total central charge $1-
3Q^2 + c_{CFT}$ is equal to 26.\foot{We have, for definiteness, taken
$X^0$ to be a time-like coordinate, but our analysis goes through without
any change even if we choose $X^0$ to be space-like.
Note that if we take $X^0$ to be space-like, and CFT to be one of the
minimal models, the system describes minimal model coupled to gravity,
with $X^0$ identified as the Liouville field.}
This describes a consistent background for the formulation of string field
theory.
In this case, we can get another consistent background for the formulation
of string field theory, by taking the direct sum of the conformal field
theory CFT$'$ and the free scalar field theory with background charge
$Q'$, so that the total central charge $1-3(Q')^2+c'_{CFT}$ again becomes
equal to 26.
We now ask the question: if we construct a string field theory in the
background of the first conformal field theory, can we construct a
classical solution of this string field theory that represents the second
conformal field theory?

Construction of the solution proceeds exactly in the same way as in the
previous case.
We start with a trial solution of the form:
$$
|\pcl\r = \lambda c_0^- c_1\bc_1 |\vp\r_{matter} \otimes |0\r_{ghost}
+\sum_{n=2}^\infty \lambda^n|\chi_n\r
\eqn\ethirtyone
$$
and try to show the existence of the solution of the equations of motion
by using the method of induction.
In this analysis we treat $\lambda$ to be of the same order as $y$ since
from hindsight we know that this would be the case.
As before, the only possible obstruction comes from the inner product of
the equations of motion with the physical states.
The assumption that the operator product of $\vp$ with itself does not
contain
any other (nearly) marginal operator except itself, implies that the only
possible obstruction comes from taking the inner product of the equations
of motion with the state $c_0^- c_1\bc_1 |\vp\r_{matter} \otimes
|0\r_{ghost}$.\foot{This state is not actually a physical state, since
$\vp$ does not have dimension (1,1). It turns out, however, that in
obtaining the solution in power series in $\lambda$, we need to treat this
state on the same footing as a physical state, otherwise in the process of
obtaining the solution we need to invert a matrix, one of whose
eigenvalues is proportional to $y$.
As a result, the inverse of this matrix, acting on an order $y^{n+1}$
term, will give an order $y^n$ term, thus upsetting the counting of powers
of $y$\SYCL.}
To order $\lambda^2$, the equation looks like:
$$
2\lambda y - {g\over\sqrt 2} C_{\vp\vp\vp}\lambda^2 + {\rm ~Polynomials~
in~\lambda~and}~ y~{\rm of~degree~}\ge 3 = 0
\eqn\ethirtytwo
$$
The above equation has a perturbative solution of the form:
$$
\lambda = {\sqrt 2\over g} {2y\over C_{\vp\vp\vp}} + \sum_{n\ge 2} a_n y^n
\eqn\ethirtythree
$$
Comparing with eq.\etwentynine\ we see that the coefficient of
perturbation $\lo$ in the conformal field theory is related to $\lambda$
through the relation:
$$
\lo = {g\over\sqrt 2}\lambda + O(\lambda^2)
\eqn\ethirtythree
$$

We can now construct the solution to any arbitrary order in $\lambda$
following the procedure outlined above (for details see ref.\SYCL).
In particular we can study the order $\lambda^3$ contribution to the
solution, and ask if we can see the change in the background charge from
$Q$ to $Q'$.
(Note that $Q^2-Q'^2$ is given by  to $(c_{CFT} - c'_{CFT})/3 =
(8 y^3/C^2_{\vp\vp\vp}) + O(y^4)$.)
For this, we note that the presence of the background charge in the $X^0$
system is equivalent to giving the dilaton a background value proportional
to $QX^0$\ANTO.
Thus a change in $Q$ will correspond to a change in the background value
of the dilaton field.
Examining the solution to order $\lambda^3$ we find that such a background
is indeed present in the solution.
In particular, we can calculate $Q-Q'$ from the solution and find
that\SYCL:
$$
Q - Q' = {8 y^3\over 6 Q (C_{\vp\vp\vp})^2}
\eqn\ethirtyfour
$$
in agreement with eq.\ethirty\ to this order.
Note that if we did not know the central charge of the perturbed conformal
field theory, the above procedure would give us a way to calculate this.

This finishes our discussion of explicit construction of the solutions of
string field theory.
We can now ask: given these solutions, what evidence do
we have that they indeed represent string theory formulated in the
background of perturbed conformal field theory?
In order to answer this question, we need to compare string field theory
action $\hs(\hp)$
formulated around these new backgrounds with string field theory action
$\ts(\tp)$ formulated directly in the background of the perturbed
conformal field theory.
First we can compare the coefficient $\hqb$ of the quadratic term in $\hs$
with the BRST charge $\tqb$ of the perturbed conformal field theory.
It turns out that to order $\lambda$ these two charges are related to each
other by an inner product preserving similarity transformation\BACKGROUND,
provided we identify $\lo_i$ with $g\lambda_i/\sqrt 2$ as in
eq.\ethirtythree.
Hence the spectrum of $\hqb$ and $\tqb$ are identical, and the kinetic
term of $\hs$ is identical to the kinetic term of $\ts$ after a linear
field redefinition.
The next question is whether the interaction terms of $\hs$ and $\ts$ can
also be shown to be identical.
In this case it turns out that to order $\lambda$, if we calculate the
S-matrix elements from the action $\hs$, they are identical to the
S-matrix elements calculated in string theory in the background of
perturbed conformal field theory\BACKTWO.\foot{This has been shown for
S-matrix elements with arbitrary number of tachyonic external legs, and
also for S-matrix elements with three arbitrary external legs.}
This, in turn, gives a strong indication that there is, in fact, a field
redefinition which takes us from $\hs(\hp)$ to $\ts(\tp)$.
Generalization of these results to higher orders in $\lambda$ has not been
carried out.
Also, so far, the explicit field redefinition which takes us from $\hs$ to
$\ts$ has not been constructed.

To summarise, we have shown how we can use string field theory to
demonstrate that conformal field theories which are related by marginal
and nearly marginal operators can be regarded as different classical
solutions of the same underlying string field theory.
We have also discussed how to construct explicit classical solutions of
string field theory representing these perturbed theories.
In this process, we have obtained string field theoretic expression for
the $\beta$-function and the central charge of the perturbed conformal
field theory.
Finally, for $c=1$ conformal field theory coupled to gravity,  we have
explicitly constructed an asymptotically flat, static (possibly singular)
solution labelled by infinite number of parameters.
For a specific set of values of the parameters the solution reduces to the
recently discovered black hole solution.

\ack
I would like to thank the organisers of the conference for their
hospitality during my stay at Stony Brook.

\refout
\end